\documentclass[aps,prb,twocolumn,superscriptaddress,floatfix,showpacs,10pt,letterpaper]{revtex4}

\usepackage{graphicx}
\usepackage{dcolumn}
\usepackage{bm}
\usepackage{amsfonts}
\usepackage{amsmath}

\begin{document}

\begin{titlepage}

\title{Tensor product representation of topological ordered phase:\\
necessary symmetry conditions}

\author{Xie Chen}
\affiliation{Department of Physics, Massachusetts Institute of Technology, Cambridge, Massachusetts 02139, USA}

\author{Bei Zeng}
\affiliation{Institute for Quantum Computing, University of Waterloo, Waterloo, Ontario, Canada}
\affiliation{Department of Combinatorics and Optimization,University of Waterloo,Waterloo,Ontario,Canada}

\author{Zheng-Cheng Gu}
\affiliation{Kavli Institute for Theoretical Physics, University of California, Santa Barbara, CA 93106, USA}

\author{Isaac L. Chuang}
\affiliation{Department of Physics, Massachusetts Institute of Technology, Cambridge, Massachusetts 02139, USA}

\author{Xiao-Gang Wen}
\affiliation{Department of Physics, Massachusetts Institute of Technology, Cambridge, Massachusetts 02139, USA}

\begin{abstract}
The tensor product representation of quantum states leads to
a promising variational approach to study quantum phase and
quantum phase transitions, especially topological ordered
phases which are impossible to handle with conventional
methods due to their long range entanglement. However, an
important issue arises when we use tensor product states
(TPS) as variational states to find the ground state of a
Hamiltonian: can arbitrary variations in the tensors that represent
ground state of a Hamiltonian be induced by local perturbations to the Hamiltonian?
Starting from a tensor product state which is the exact
ground state of a Hamiltonian with $\mathbb{Z}_2$
topological order, we show that, surprisingly, not all
variations of the tensors correspond to the variation of the
ground state caused by local perturbations of the
Hamiltonian. Even in the absence of any symmetry requirement
of the perturbed Hamiltonian, one necessary condition for
the variations of the tensors to be physical is that they
respect certain $\mathbb{Z}_2$ symmetry. We support this claim
by calculating explicitly the change in topological
entanglement entropy with different variations in the
tensors. This finding will provide important guidance to
numerical variational study of topological phase and phase
transitions. It is also a crucial step in using TPS to study
universal properties of a quantum phase and its topological
order.
\end{abstract}


\maketitle

\vspace{2mm}

\end{titlepage}

\section{Introduction}

The central task in the study of quantum many-body systems
is the classification of possible phases of matter and the
understanding of phase transitions between them. Of
particular interest is the study of systems at zero
temperature, where a whole new realm of quantum effects
emerge, and that is what we will focus on in this paper.
Landau's general principle for understanding 
continuous phase transitions based on symmetry breaking and
local order parameter \cite{Landau} does not apply to all
phases and phase transitions.  Topological order,\cite{Wrig,FQH_TO} in
particular, is not related to any symmetry properties and topological
phase transitions may happen between systems with the same
\cite{Wen_93,RG,Wen_2000,Wqosl} or incompatible
\cite{Senthil} symmetries. It has been the subject of
intensive research and will be the central topic of this
paper. Aside from the lack of a qualitative understanding,
what makes the problem harder is the fact that for quantum
systems whose components strongly interact with each other,
direct numerical approach is limited due to strong
entanglement among the particles. Generically, the space
required for description of a quantum system grows
exponentially with system size, hence limiting direct
numerical simulations to systems usually too small for any
practical purpose.

Recently, insights into quantum many-body systems from both
condensed matter physics and quantum information science
have led to the discovery of the tensor product
representation of quantum states (also called the projected
entangled pair
states)\cite{GMN0391,NMG0415,M0460,VC0466,TPS_review}, which
provides a promising variational approach to study zero
temperature quantum phase and phase transitions.
Representing quantum many-body states with a network of
tensors, tensor product states (TPS) are proven to be
efficient for the study of one dimensional quantum
systems\cite{Verstraete_06, Schuch_1D, Aharonov}. The higher
dimensional generalization of this approach may not be as
efficient, yet study has shown that it reproduces many known
results and may reveal new features in systems not solvable
in any conventional way\cite{Murg,TERG,JWX}. The strength of
the approach lies in the fact TPS can describe long range
entanglement that are present in a large class of
topologically ordered states.\cite{TPS_TO_Wen,TPS_TO_Vidal}
So the variational approach based on TPS can include both
topologically ordered states and symmetry breaking states
and can produce a phase diagram that contains both types of
states.  In contrast, the conventional mean-field/variational
approaches are based on states with no long range
entanglement, which exclude the topologically ordered
states from the very beginning.  We also note that the
entanglement of a simple TPS satisfies an area
law\cite{PEPS}, which coincides with the scaling of
entanglement in the ground state of most known non-critical
systems\cite{area_law}.

In the variational approach based on TPS, we try to find a TPS
which minimizes the average energy of a local Hamiltonian. As we
change the Hamiltonian by adding perturbation, the tensors
in the TPS are also changed in order to minimize the average
energy for the new Hamiltonian. While local physical perturbations 
can always be reflected by variations in the tensors, the other
direction of this problem remains unclear:
can an arbitrary variation of the tensor be induced by a
local perturbation of the Hamiltonian?

This is a very important question if we want to discuss
phase based on states.  Because phase is defined as a region
in Hamiltonian space, where any two points $H_1$, $H_2$
within the region can be connected by a smooth path without
encounter singularities ({\it i.e.} phase transitions).  So
the question about tensor and phase becomes, which set of
states in Hilbert space correspond to such a region in
Hamiltonian space and which set of tensors in the tensor
space represent these states.  Starting from one point in
the phase region, we would like to know what kind of
variations in the tensors correspond to local
perturbations to the Hamiltonian. 

We can discuss this important question in more concrete
setting. Assume a TPS $\Psi_T$ minimizing the average energy
of a Hamiltonian $H$ has a property.  We would like to
ask if the property is a universal property of a phase, or
just a special property of $H$. If the property is a
universal property of a phase, then the ground state
$\Psi_T+\Delta \Psi_T$ for the perturbed Hamiltonian
$H+\Delta H$ still has the same property.  If the property is a
special property of $H$, the ground state for the
perturbed Hamiltonian will lose this property. It is the
collection of universal properties that defines a phase.  So
a study of universal properties is a study of phases.  If
all the variations of the tensors can be induced by local
perturbations of the Hamiltonian, then we can study the
stability of a property against local perturbation $\Delta
H$ of the Hamiltonian by studying the stability of a property
against variations of the tensors. This will give us a
powerful tool to study phases using TPS. 

Unfortunately, it turns out that not all variations of the tensors can be induced by local perturbations of the Hamiltonian, as we show in this paper. For a generic TPS, which satisfies a condition called injectivity \cite{PEPS_H}, tensor variations indeed correspond to Hamiltonian perturbations. However, this is not true in the general case, as we show in this paper with a special system with topological order. So it is not easy to study
universal properties and phases using TPS. In order to use
TPS to study phases and phases transitions, we need to find
the subset of variations in tensors that are physical, {\it i.e.}
corresponding to local perturbations of the Hamiltonian.

For clarity, we will always refer to small changes in the
Hamiltonian as `perturbation' and to those in the tensors as
`variation'.  Without any efficient method to solve for
exact TPS representation of ground states of quantum
many-body systems, finding the subset of the variations of
the tensors that can be induced by local perturbation of
Hamiltonian is in general very difficult.

We want to, in particular, study this problem for topologically ordered phases.
As TPS can give a simple description of a large class of topological ordered states, we expect that it might provide a powerful tool for studying topological phases in general. As we know, topologically ordered phases are proven to be stable against
any local perturbations of the
Hamiltonian\cite{FQH_TO,Hastings_Qcon,Hastings_TO}. That is,
the topological properties, such as ground state
degeneracy\cite{FQH_TO} and  quasi-particle
statistics\cite{Arovas, Kitaev}, are robust under any local
perturbation to the Hamiltonian.  So in the TPS approach to
topologically ordered phase, it is natural to ask: are those
topological properties robust against any variation of
tensors, that is, for any tensor which represents a
topologically ordered state, is the topological order robust
against arbitrary variation in the tensors? Surprisingly, we find
that this is not true.

We focus on the $\mathbb{Z}_2$ topological order represented
by an ideal TPS in this paper and study how the topological
order of the state changes as we vary certain parameters in
the representing tensors. We characterize topological order by calculating the topological entanglement entropy $S_{tp}$ \cite{Stp_Kitaev,Stp_Wen} for the state and observe 
that topological order ({\it i.e.} the topological
entanglement entropy $S_{tp}$) is stable only against
variations of the tensors that preserve certain
$\mathbb{Z}_2$ symmetry of the tensors.  Since the
topological order is robust against any local
perturbations of Hamiltonian, this result shows that not all
variations of the tensors correspond to local perturbations
of Hamiltonian.  For this $\mathbb{Z}_2$ model, we show that in the generic case $\mathbb{Z}_2$ symmetry is a necessary
condition for the variations in tensors to correspond to
physical perturbations of the Hamiltonian. This claim is
further supported by checking stability of the topological
Renyi entropy of TPS with $\mathbb{Z}_2$ symmetry preserving
variations and $\mathbb{Z}_2$ symmetry breaking variations of the tensors
respectively.

While calculating $S_{tp}$ for a general state is exponentially
hard\cite{Hamma}, we find efficient ways to do so for two sets of TPS
near the ideal TPS with $\mathbb{Z}_2$ topological order.
For a general TPS, we calculate the topological Renyi entropy by mapping it to the contraction of a 2D tensor network, which is accomplished by using the Tensor Entanglement Renormalization Algorithm\cite{Nave}. Hence we are able to calculate topological entropy for regions much larger than was possible previously and determine the topological order of the state more accurately.

Our result on the stability of topological order will help us in the TPS based variational
approach to $\mathbb{Z}_2$ topological phase: we should only
consider the variations of the tensors within the subspace
of tensors with $\mathbb{Z}_2$ symmetry. The
$\mathbb{Z}_2$ symmetry condition and possibly other
conditions will help us to understand the physical
variations of tensors in TPS. This is crucial in using TPS
to study quantum phases and quantum phase transitions. It
may even lead to a classification of topological order.

This paper is organized as follows. We start by introducing an `ideal'
lattice spin model with $\mathbb{Z}_2$ topological order and show
how the presence of topological order in the ground state
wave function can be understood nicely with a physical mechanism
called `string-net condensation'. Such a physical picture naturally
gives rise to a simple tensor product representation of the
wave function, to which we then add two kinds of local variations,
`string tension' and `end of strings'. By calculating topological
entanglement entropy numerically for the first case and analytically
for the second case, we show how topological order is stable against
$\mathbb{Z}_2$ preserving variations (`string tension'), but breaks
down immediately when $\mathbb{Z}_2$ symmetry is broken (by `end of
strings'). We then randomly picked ~200 tensors in the neighborhood
of the ideal $\mathbb{Z}_2$ TPS and calculate the topological Renyi
entropy of the corresponding states. Tensors with and without
$\mathbb{Z}_2$ symmetry demonstrate totally different behavior as
system size scales up. We discuss in the last section the
implications of our findings in variational studies of topological
phase and phase transitions. The details of the calculations are
given in the appendix.

\section{Models and Result}

\subsection{Spin model with $\mathbb{Z}_2$ topological order}

We start from an exactly solvable model which has
$\mathbb{Z}_2$ topological order\cite{RS9173,W9164,Qdouble}.
In this section, we give the system Hamiltonian, find the ground state
wave function and explain its structure and how that leads to
a nontrivial topological order which can be detected with
topological entanglement entropy. With these insights about
the state we then present a simple tensor product
representation of this wave function. 

The model is defined on a two-dimensional hexagonal lattice
where each link is occupied by a qubit (spin-$1/2$).
The Hamiltonian is a sum of commuting projection operators
\begin{equation}
H^0_{\mathbb{Z}_2} = -\sum_{p} \prod_{i \in p} X_i -\sum_{v}
\prod_{j \in v} Z_j \label{H_Z2_0}
\end{equation}
$X$ and $Z$ are qubit Pauli operators defined as
\begin{math}
X = \left(\begin{array}{cc}
0 & 1\\
1 & 0
\end{array}\right),
Z = \left(\begin{array}{cc}
1 & 0\\
0 & -1
\end{array}\right)
\end{math}.
$p$ stands for each hexagon plaquette in the lattice and $\prod_{i
\in p} X_i$ is the tensor product of six $X$ operators around the
plaquette. $v$ stands for each vertex and $\prod_{j \in v} Z_j$ is
the tensor product of three $Z$ operators connected to the vertex.
The ground state wave function has a nice interpretation using the
`string-net' picture where state $|0\rangle$ corresponds to no
string on a link and state $|1\rangle$ corresponds to the presence
of a string. The vertex term $\prod_{j \in v} Z_j$ enforces that there
are even number of strings connected to each vertex and hence the
strings form closed loop while the plaquette term $\prod_{i \in p}
X_i$ gives dynamics to the closed loops. The ground state
wave function is an equal weight superposition of all closed loop
configurations on the lattice.
\begin{equation}
|\Phi_{\mathbb{Z}_2}\rangle = \sum_{cl} |\phi_{cl}\rangle \label{phi_z2}
\end{equation}
The normalization factor is omitted. If we refer to each closed loop
configuration as a string-net, the appearance of $\mathbb{Z}_2$
topological order in this system has then a natural interpretation
as being due to the condensation of string-nets. We will refer to this
model as the ideal $\mathbb{Z}_2$ model.

\begin{figure}[htb!]
\centering
\includegraphics[width=2.5in]{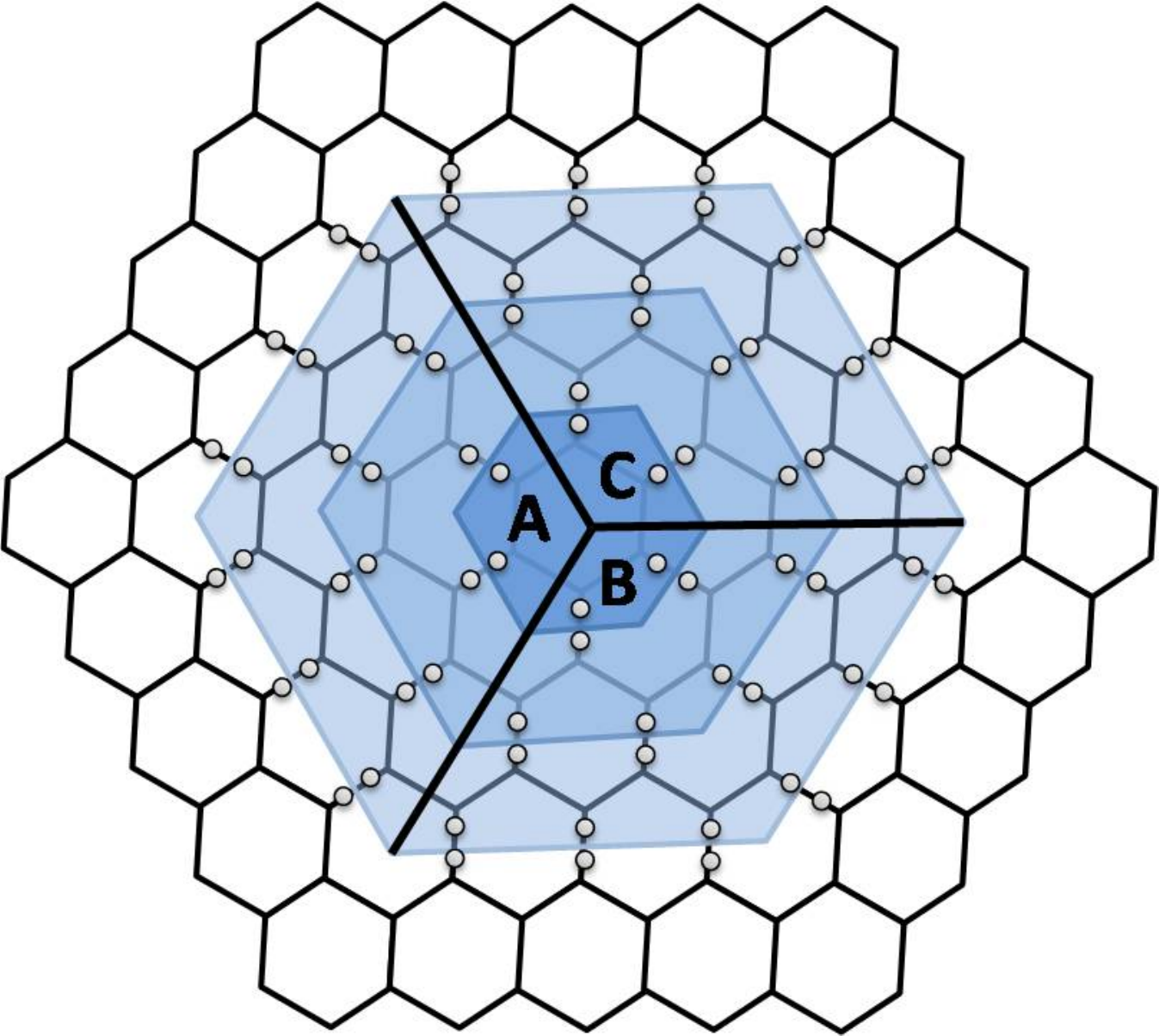}
\caption{Hexagonal lattice where each link is occupied by two
qubits. Ideal $\mathbb{Z}_2$ model can be defined on this lattice
and the red, yellow, blue sets of qubits correspond respectively to
the plaquette, vertex, and link terms in the $\mathbb{Z}_2$
Hamiltonian.} \label{fig:hex}
\end{figure}

For simplicity of discussion, we split each qubit on a link into two
qubits as illustrated in Fig \ref{fig:hex}. The string-net condensed
$\mathbb{Z}_2$ wave function on the original lattice can be naturally
extended to a state on the new lattice by replacing a $0$ link with
$00$ and a $1$ link with $11$. This new state is still an equal
weight superposition of all string-nets and hence maintains the
$\mathbb{Z}_2$ topological order. The new system Hamiltonian can be
obtained from the old one by adding a $-ZZ$ term to each link and
expand the plaquette term into a product of $X$ operators on all
twelve qubits around the plaquette
\begin{equation}
H_{\mathbb{Z}_2} = -\sum_{p} \prod_{i \in p} X_i -\sum_{v} \prod_{j
\in v} Z_j - \sum_{l} Z_{l_1}Z_{l_2} \label{H_Z2}
\end{equation}
where $l$ denotes all the links and $l_1,l_2$ are the two qubits on
the link. It is easy to see that the new Hamiltonian indeed has the
new string-net condensed state as its ground state. The topological
order of the system can be detected from the ground state
wave function by calculating the topological entanglement entropy of
the state. The mapping to the new lattice allows this
calculation to be carried out exactly in a few steps, as
illustrated below.

According to the definition of topological entanglement entropy in Ref.
\onlinecite{Stp_Kitaev}(or equivalently defined in Ref. \onlinecite{Stp_Wen}), we
take out a simply connected region from the whole lattice and divide
it into three parts $A$, $B$, $C$ as shown in Fig \ref{fig:hex}. By
calculating the entanglement entropy for regions $A$, $B$, $C$,
$AB$, $AC$, $BC$, $ABC$ and combining them according to
\begin{equation}
S_{tp}= S_A + S_B + S_C - S_{AB} - S_{BC} - S_{AC} + S_{ABC} \label{Stp}
\end{equation}
we arrive at the topological entanglement entropy $S_{tp}$
of the state. The above definition needs to be applied to
regions much larger than the correlation length of the
state. For the state in consideration, the correlation
length is zero and the calculation gives the right result
for whatever regions we take. We divide the regions by
cutting through the pair of qubits on boundary links as
illustrated in Fig \ref{fig:hex}. For a region with $n$
outgoing links on the boundary, there are $2^{n-1}$
orthogonal boundary configurations due to the closed loop
constraint of the wave function. Tracing out each boundary
configuration contributes equally and independently to the
entropy of the region and hence $S = n-1$, which includes
one term proportional to the length of the boundary $n$ and
one constant term $-1$. The combination in the definition of
$S_{tp}$ makes sure that the boundary terms of different
regions cancel out with each other, so topological
entanglement entropy for the state is then $S_{tp}=1$.

\begin{figure}[htb!]
\centering
\includegraphics[width=3in]{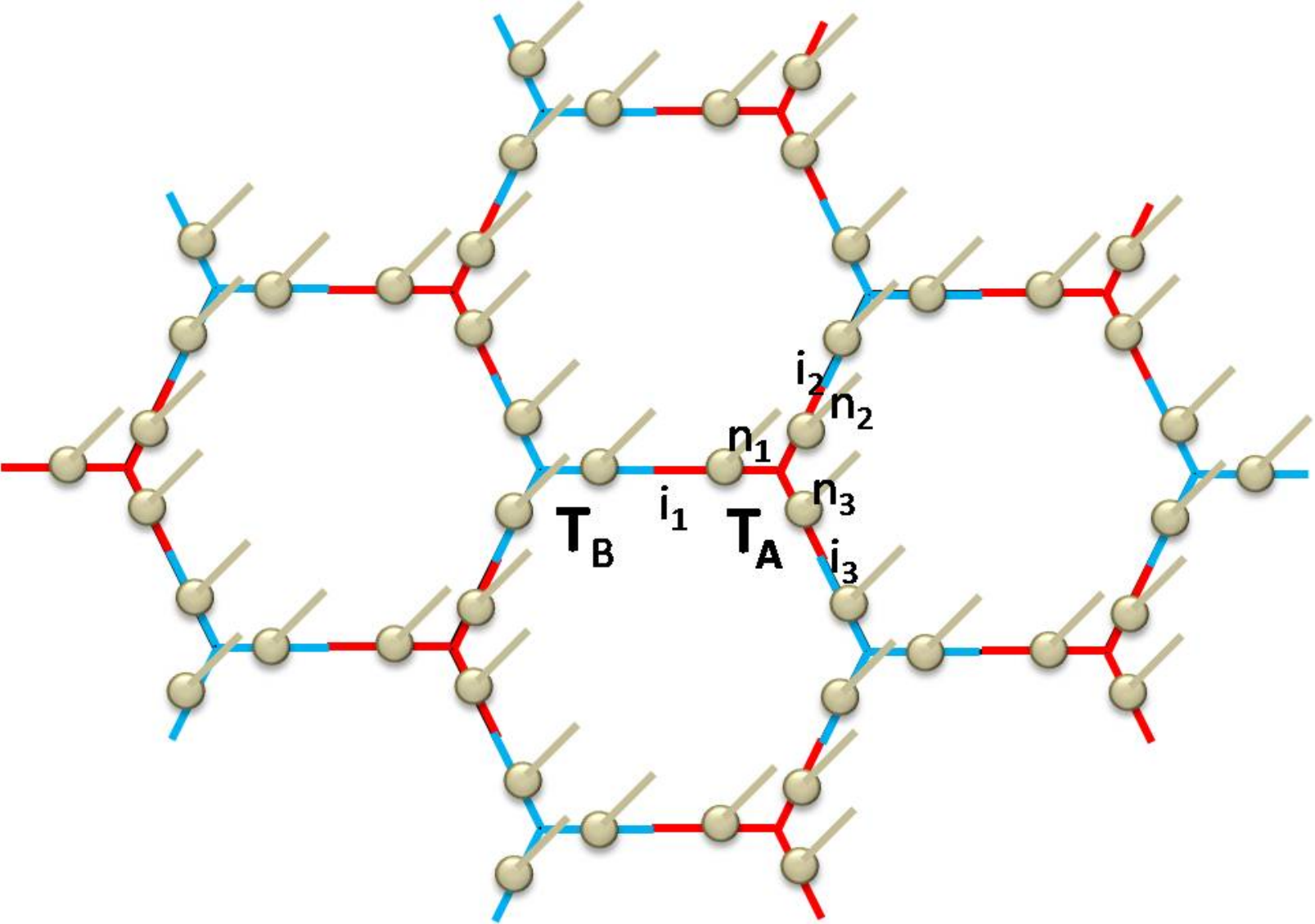}
\caption{Tensor product representation of $\mathbb{Z}_2$ ground state.
One tensor is assigned to every three qubits connected to the same vertex.
Red tensors are on vertices in sublattice $A$ and blue ones are in sublattice $B$.
The out-of-plane gray links represent the physical indices of the qubits.
The tensors connect according to the underlying hexagonal lattice.
} \label{fig:TPS}
\end{figure}

This globally entangled state has yet a surprisingly simple local
representation using the tensor product language. A tensor product
state of two dimensional lattice model is represented by associating
with each lattice site $m$ a set of $s$ tensors $T^{m}_{[k]}(\alpha\beta\gamma...)$,
$k = 1,2, ..., s$, where $s$ is the dimension of local Hilbert
space at site $m$. $k$ is called the physical index of the tensor. $\alpha\beta\gamma$, the inner indices of the tensors, connect to each other and form a
graph. The wave function (unnormalized) is then given by
\begin{equation}
|\psi\rangle = \sum_{k_1,k_2,...k_m...}\mathcal{C}(T^{1}_{[k_1]}T^{2}_{[k_2]}...T^{m}_{[k_m]}...)|k_1 k_2 ... k_m...\rangle \label{TPS}
\end{equation}
where $\mathcal{C}$ denotes tensor contraction of the inner indices according to
the connection graph. We omit the inner indices here. (We will in most cases ignore
normalization of wave function in the following discussion
and mention specifically when normalization is needed.) The
tensors representing the ground state of the ideal
$\mathbb{Z}_2$ model can be given as follows.  We group
every three qubits connected to the same vertex together and
assign a rank three tensor to each of the eight physical
basis states of the three qubits. Now every physical index $k$ in Eqn.\ref{TPS} is represented with three bits $n_1n_2n_3$. 
The eight tensors are:
\begin{equation}
\begin{array}{ll}
T_{[000]}(000) = 1 & T_{[011]}(011) = 1 \\
T_{[101]}(101) = 1 & T_{[110]}(110) = 1 \\
\text{all other terms are zero}
\end{array}
\label{TPS_Z2}
\end{equation}
The physical indices $[n_1n_2n_3]$ correspond to the out-of-plane gray
links in Fig \ref{fig:TPS}. The inner indices $(i_1i_2i_3)$ correspond to the
red/blue links in Fig \ref{fig:TPS}. The inner and physical
indices all have dimension two and are given in the same
order as shown in Fig \ref{fig:TPS}. Hence the inner indices
truthfully reflect the configuration of the physical space
and only configurations with even number of strings at each
vertex are allowed. It can then be checked that only
string-net configurations have non-zero amplitude in this
representation and the amplitude are actually all equal.
Therefore, the tensors given in \ref{TPS_Z2} indeed
represent a string-net condensed state--the ideal
$\mathbb{Z}_2$ ground state.

This set of tensors serves as a starting point for our
variational study of topological phase transitions and we
wish to know what kind of variations of the tensors
correspond to physical perturbations of the Hamiltonian. We
will study first two specific cases in the following two
sections.

\subsection{$\mathbb{Z}_2$ model with string tension}

Suppose that we want to know how magnetic field in the $Z$
direction might affect topological order. The perturbed
Hamiltonian reads:
\begin{eqnarray}
H & = & H_{\mathbb{Z}_2} + \lambda\sum_k Z_k \\
  & = & -\sum_{p}\prod_{i \in p} X_i -\sum_{v}\prod_{j
\in v} Z_j - \sum_{l}Z_{l_1}Z_{l_2} + \lambda\sum_k Z_k \nonumber
\label{Z2+Z}
\end{eqnarray}
The $Z_k$ term commute with the vertex and link term
$\prod_{j \in v} Z_j$, $Z_{l_1}Z_{l_2}$ in the unperturbed
Hamiltonian, so the closed loop constraint is maintained.
The ground state wave function is still a superposition of
string-net configurations, but with different weight. The
magnetic field adds energy to each string segment, therefore
one reasonable guess about the ground state is that each
string-net configuration has weight exponential in its total
length of string.  \begin{equation}
|\Phi^g_{\mathbb{Z}_2}\rangle = \sum_{cl}
g^{-L(\phi_{cl})/2}|\phi_{cl}\rangle \label{Phi_Z2_g}
\end{equation}
where the summation is over all string-net configurations
and $L(\phi_{cl})$ is the total string length of a
configuration.  This weighted superposition can still have a
simple tensor product representation by locally modifying
the tensors in Eq.\ref{TPS_Z2} to \begin{equation}
\begin{array}{ll} T_{[000]}(000) = g & T_{[011]}(011) = 1 \\
T_{[101]}(101) = 1 & T_{[110]}(110) = 1 \\ \text{all other
terms are zero}
\end{array}
\label{TPS_Z2_g}
\end{equation}
For $g>1$, the weight of each string segment is smaller by a
factor of $g^{-1/2}$ than that of no string, lowering the
weight of string-net configurations exponentially.
Physically, we can imagine this is due to some kind of
tension in the strings. Therefore, we refer to this
wave function as $\mathbb{Z}_2$ state with string
tension ($g$). This state cannot be the exact ground state of
the Hamiltonian given in Eq.\ref{Z2+Z}, but it is possible
that it gives a qualitatively right and quantitatively close
approximation to the ground state and hence might be a good
guess for variational study. One necessary condition for
this conjecture to be true is that the topological order of
the state remains stable with $g$ close to $1$, and this is
indeed the case as we will show below by calculating
topological entanglement entropy of the state.

\begin{figure}[htp] \centering
\includegraphics[width=3.5in]{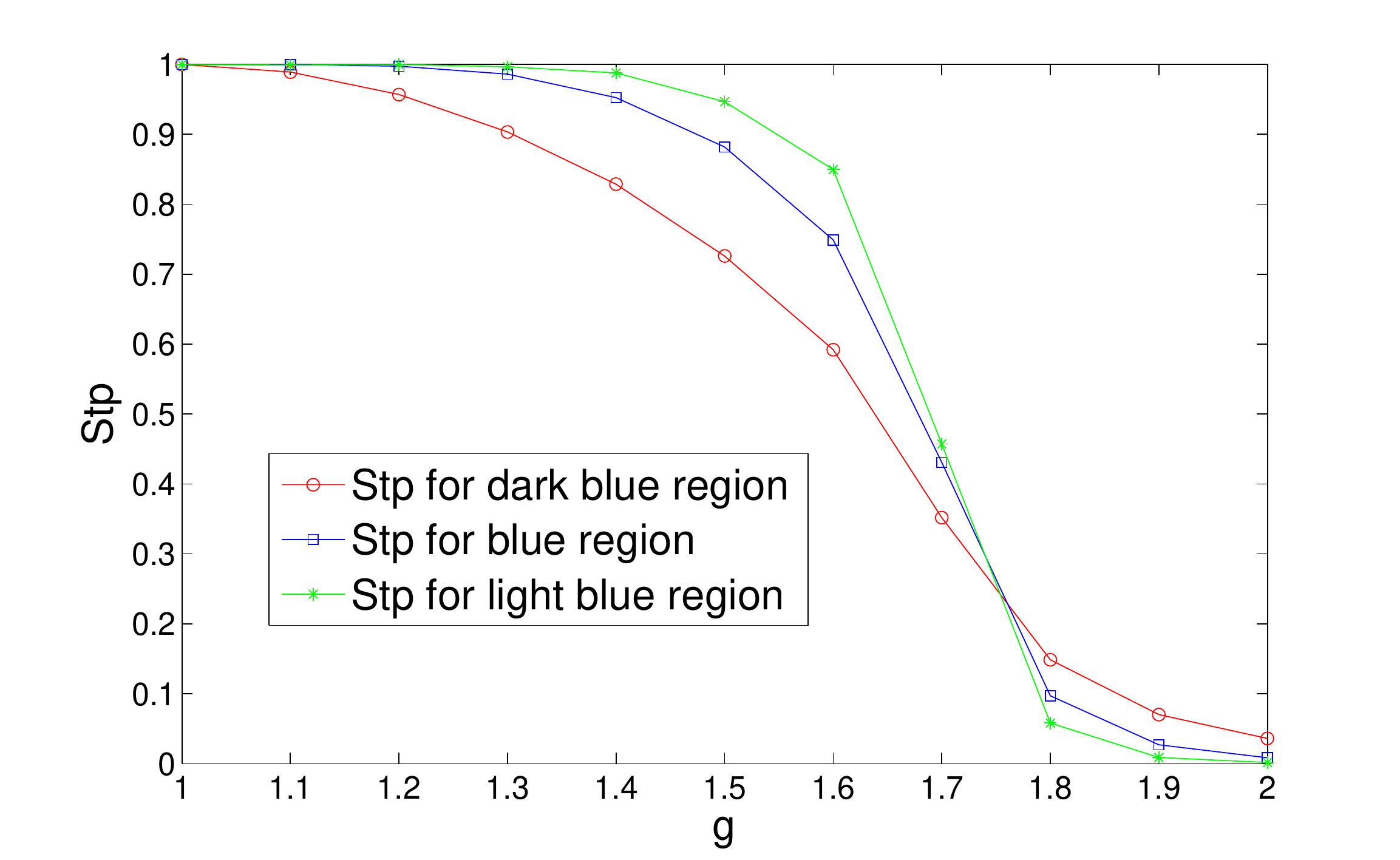} \caption{Topological
entanglement entropy $S_{tp}$ of the $\mathbb{Z}_2$ model
with string tension calculated for the three blue regions as
shown in Fig \ref{fig:hex}. $1/g$ is the weight of each
string segment relative to vacuum. $S_{tp}$ remains stable
for a finite region away from the ideal $\mathbb{Z}_2$ TPS
and drops sharply to zero at $g \sim 1.75$.}
\label{fig:Stp_g}
\end{figure}

In general, this computation is intractable. The equality in
Eq.\ref{Stp} holds only in the limit of infinitely large
regions $A$, $B$, $C$. Therefore the computation involves
diagonalization of exponentially large matrices, each
element of which takes exponential time to calculate. For
$\mathbb{Z}_2$ state with string tension, we circumvent this
difficulty by appealing to the special structure of the
tensors in Eq.\ref{TPS_Z2_g}.  In order to find the
entanglement entropy of a region, we map the computation of
the eigenvalues of the reduced density matrix to the
contraction of a two dimensional tensor network. While
contracting general two-dimensional tensor networks is $\#P$
hard, the tensor networks we are dealing with turn out to be
of a special type, called the `matchgate' tensor \cite{matchgate}.
`Matchgate' tensors can be contracted efficiently, which
leads to an efficient algorithm for determining the
topological entanglement entropy of this state. The details
of the procedure will be explained in appendix A.

In the computation, we take the total system size to be
large enough such that it does not affect the result of the
computation. Taking the size of regions $A$, $B$ and $C$ to
infinity is hard. We manage to carry out the computation for
progressively larger regions as shown in Fig \ref{fig:hex},
with $18$, $72$ and $162$ qubits inside respectively. The
resulting topological entanglement entropy is plotted in Fig
\ref{fig:Stp_g}. We do see a trend of sharper drop of
$S_{tp}$ from $1$ to $0$ as the size of the region is
increased. At $g$ close to zero or $g$ very large when the
correlation length is small compared to the size of the
region, the calculated value for $S_{tp}$ is reliable and we
find that it remains stable within a finite range of the
ideal $\mathbb{Z}_2$ TPS and drops to zero beyond certain
critical string tension $g_c$. We can decide from the plot
the critical point $g_c$ to be around $1.75$. We would like
to comment that the tensor network representing the norm of
the $\mathbb{Z}_2$ state with string tension $g$ is the same
as that representing the partition function of classical
Ising model on triangular lattice with coupling constant $J$
at inverse temperature $\beta = \ln g / (2J)$. A phase
transition at $g = \sqrt{3}$ is known for this classical
model. Our calculation for the quantum model confirms this
observation and shows that the quantum phase transition is
in fact topological.

The stability of topological order at $g\ge 1$ is a necessary
condition for string tension to correspond to local
Hamiltonian perturbations. In this particular case, we can
actually find the corresponding perturbations explicitly.
The relation we mentioned above between the $\mathbb{Z}_2$
state with string tension and 2D classical Ising model at
finite temperature allows the construction of a continuous
family of parent Hamiltonian $H(g)$ for the quantum
states\cite{PEPS}. The Hamiltonian $H(g)$ is local, has the
state $|\Phi^g_{\mathbb{Z}_2}\rangle$ as its exact ground
state and remains gapped for $g<g_c$. Therefore we can
conclude that string tension can be induced by local
perturbations of the Hamiltonian and hence is an allowed
variation of the $\mathbb{Z}_2$ tensors.

\subsection{$\mathbb{Z}_2$ model with end of strings}

Another simple model one might want to study is the
$\mathbb{Z}_2$ model with magnetic field perturbation in the
$X$ direction.  \begin{eqnarray} H & = & H_{\mathbb{Z}_2} +
\lambda\sum_k X_k\\ & = & -\sum_{p}\prod_{i \in p} X_i
-\sum_{v}\prod_{j \in v} Z_j - \sum_{l}Z_{l_1}Z_{l_2} +
\lambda\sum_k X_k \nonumber \label{Z2+X}
\end{eqnarray}
The action of the perturbation operator $X_k$ on
$\mathbb{Z}_2$ ground state will flip a link from no string
to having a string (or back) and hence break one or more
closed loops. The perturbed ground state would contain
configurations with end of strings. In tensor language, this
corresponds to allowing some odd configurations to be
non-zero. Taking the translational and rotational symmetries
of the Hamiltonian into consideration, one might expect that
the following tensors which assign a small and equal weight
$\epsilon$ to all odd configurations might represent a good
trial wave function for the ground state.  \begin{equation}
\begin{array}{ll} T_{[000]}(000) = 1 & T_{[011]}(011) = 1 \\
T_{[101]}(101) = 1 & T_{[110]}(110) = 1 \\ T_{[001]}(001) =
\epsilon & T_{[010]}(010) = \epsilon \\ T_{[100]}(100) =
\epsilon & T_{[111]}(111) = \epsilon \\ \text{all others are
zero}
\end{array}
\label{TPS_Z2_e}
\end{equation}
Again the inner indices $(i_1i_2i_3)$ truthfully reflect the
configurations of the physical indices $[n_1n_2n_3]$. When
$\epsilon=0$, this is reduced to the tensors in the ideal
$\mathbb{Z}_2$ TPS. When $\epsilon$ is non-zero, the
wave function contains all possible string configurations,
closed loop or open string. The weight of each string
configuration is exponentially small in the number of end of
strings contained.  \begin{equation}
|\Phi^{\epsilon}_{\mathbb{Z}_2}\rangle = \sum_{sc}
\epsilon^{q(\phi_{sc})} |\phi_{sc}\rangle \label{Phi_z2_e}
\end{equation}
where the summation is over all possible string
configurations and $q(\phi_{sc})$ is the number of end of
strings in a particular configuration.

To see how topological order of the state changes as
$\epsilon$ varies from $0$, we again calculate the
topological entanglement entropy of the state. In this case,
it turned out that analytical calculation is possible. The
detailed procedure is given in appendix B. We find that for
any finite value of $\epsilon$, when system size goes to
infinity, $S_{tp}$ goes to zero. Hence topological order is
unstable under this kind of variation. At first sight this
may be a surprising result, as we are only changing the
tensors locally and we are not expected to change the global
entanglement pattern of the state. However, when we write
out the wave function explicitly we will see that we have
actually induced global changes to the state.  The
wave function in Eq. \ref{Phi_z2_e} can be expanded in powers
of $\epsilon$ as \begin{equation}
|\Phi^{\epsilon}_{\mathbb{Z}_2}\rangle =
|\Phi_{\mathbb{Z}_2}\rangle + \epsilon^2 \sum_{v_i,v_j}
|\Phi^{v_i,v_j}_{\mathbb{Z}_2}\rangle + ... \label{Phi_vivj}
\end{equation}
where the $v$'s are any vertices in the lattice.
$|\Phi^{v_i,v_j}_{\mathbb{Z}_2}\rangle$ is an excited
eigenstate of the $\mathbb{Z}_2$ Hamiltonian (Eq.
\ref{H_Z2}) which minimizes energy of all local terms except
the vertex terms at $v_i$, $v_j$ and is hence an equal
weight superposition of all configurations with end of
strings at $v_i$ and $v_j$. Note that end of strings always
appear in pairs. We will call such a pair a defect in the
string-net condensate.  $v_i$, $v_j$ can be separated by any
distance and the number of local operations needed to take
$|\Phi_{\mathbb{Z}_2} \rangle$ to
$|\Phi^{v_i,v_j}_{\mathbb{Z}_2}\rangle$ scale with this
distance.

On the other hand, with arbitrary local perturbation to the
dynamics, the Hamiltonian reads \begin{equation} H' =
H_{\mathbb{Z}_2} + \eta \sum_{u} h_u   \label{H'}
\end{equation}
where $h_u$'s are any local operator and $\eta$ is small. The
perturbed ground state wave function will contain terms like
$|\Phi^{v_i,v_j}_{\mathbb{Z}_2}\rangle$ but only with weight
$~\eta^{distance(v_i,v_j)}$. When $v_i$, $v_j$ are separated by a
global distance, the weight will be exponentially small. Hence a
constant, finite weight $\epsilon^2$ for all
$|\Phi^{v_i,v_j}_{\mathbb{Z}_2}\rangle$ as required in Eq.
\ref{Phi_vivj} is not possible. Therefore, while we are only
modifying the tensors locally, we introduce global `defects' to the
state, which cannot be the result of any local perturbation to the
Hamiltonian. We can, of course, design a Hamiltonian $H_{\epsilon}$ which has $|\Phi^{\epsilon}_{\mathbb{Z}_2}\rangle$ as its exact ground state using the method introduced in Ref. \onlinecite{PEPS_H}. However, $H_{\epsilon}$ will not be able to smoothly connect to $H_{\mathbb{Z}_2}$ as $\epsilon \to 0$. 

\subsection{Necessary symmetry condition}
The two kinds of tensor variations we have studied have
drastically different effects on the topological order of
the state. While the first type corresponds to local
perturbations of the Hamiltonian and keeps topological order
intact, the second type does not have a physical
correspondence and destroys the topological order
completely. What leads to such a difference? Given a general
variation of $\mathbb{Z}_2$ tensor, how can we tell if it is
allowed?

\begin{figure}[htp] \centering
\includegraphics[width=1.5in]{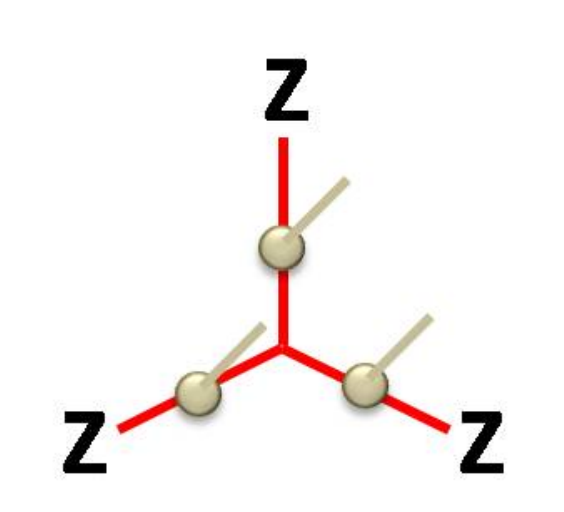} \caption{Symmetry of
the $\mathbb{Z}_2$ tensor. The tensor representing the ideal
$\mathbb{Z}_2$ state is invariant under the
action of $Z \otimes Z \otimes Z$ to its inner indices. The
variation in string tension (Eq.\ref{TPS_Z2_g}) does not
break this symmetry and topological order is stable. The
variation with end of strings (Eq.\ref{TPS_Z2_e}) breaks this
symmetry and destroys topological order} \label{fig:symm}
\end{figure}

We observe that the tensor representing the ideal $\mathbb{Z}_2$
state (Eq.\ref{TPS_Z2}) has certain inner
symmetry, that is, the tensor is invariant under some
non-trivial operations on the inner indices, as shown in
Fig\ref{fig:symm}.

$Z$ does nothing to the tensor when the index is $0$ and
changes the sign of the tensor when the index is $1$. In the
ideal $\mathbb{Z}_2$ tensor, only even configurations
of the inner indices are non-zero. Hence applying $Z$ at the
same time to all three inner indices doesn't change the
tensor. That is, $Z \otimes Z \otimes Z$ is a symmetry of
the tensor. As $Z \otimes Z \otimes Z$ squares to identity,
we will say that the tensor has $\mathbb{Z}_2$ symmetry. 
Note that we can insert a set of unitary
operators $U,U^{\dagger}$ between any connected links in a
tensor network without affecting the result of tensor
contraction and hence the quantity represented by the tensor
network. Therefore, the $\mathbb{Z}_2$ symmetry could take any 
form which is local unitary equivalent to $Z \otimes Z \otimes Z$.
This $\mathbb{Z}_2$ symmetry is closely related to the closed loop constraint of the state. 
Due to this symmetry, the tensor network cannot be `injective'
as defined in Ref. \onlinecite{PEPS_H}.

Adding string tension to the $\mathbb{Z}_2$
tensor (Eq.\ref{TPS_Z2_g}) does not violate this symmetry, as
all the odd terms of inner indices are still zero. We found
that topological order of the state is stable with small
string tension. On the other hand, adding end of
strings (Eq.\ref{TPS_Z2_e}) breaks this symmetry for any
finite $\epsilon$. In general, assume the variation of the tensor $T$ contains a $\mathbb{Z}_2$ symmetry breaking term $dT$ of magnitude $\delta$. Such a term would represent an end of string in the tensor network. To the leading order in $\delta$, the wavefunction would contain terms on the order $O(\delta^2)$ with $dT$ on two of the sites and $T$ on the others. In the physical space, this would correspond to an open string configuration(up to local unitaries at the ends). The weight of such a term is $O(\delta^2)$ even though the two sites with $dT$ may be globally apart, hence introducing global defects to the wavefunction and breaking topological order. Such defect terms cannot be created by local perturbation to the Hamiltonian. Therefore,
$\mathbb{Z}_2$ symmetry breaking variations to the tensors are not
allowed and preserving $\mathbb{Z}_2$ symmetry of the tensor is shown to be
a necessary condition for any variation of the ideal
$\mathbb{Z}_2$ tensor to be physical. This argument is valid for a generic $\mathbb{Z}_2$ breaking variation. There can be specially designed cases where $\mathbb{Z}_2$ breaking variations does not lead to breakdown of topological order, e.g. when such variations only occur within a finite region of the system or different contributions to the global defects exactly cancel each other. However, for a random $\mathbb{Z}_2$ breaking variation, topological order will be lost and it cannot correspond to local perturbation of Hamiltonian.

\begin{figure}[htp] \centering
\includegraphics[width=3.5in]{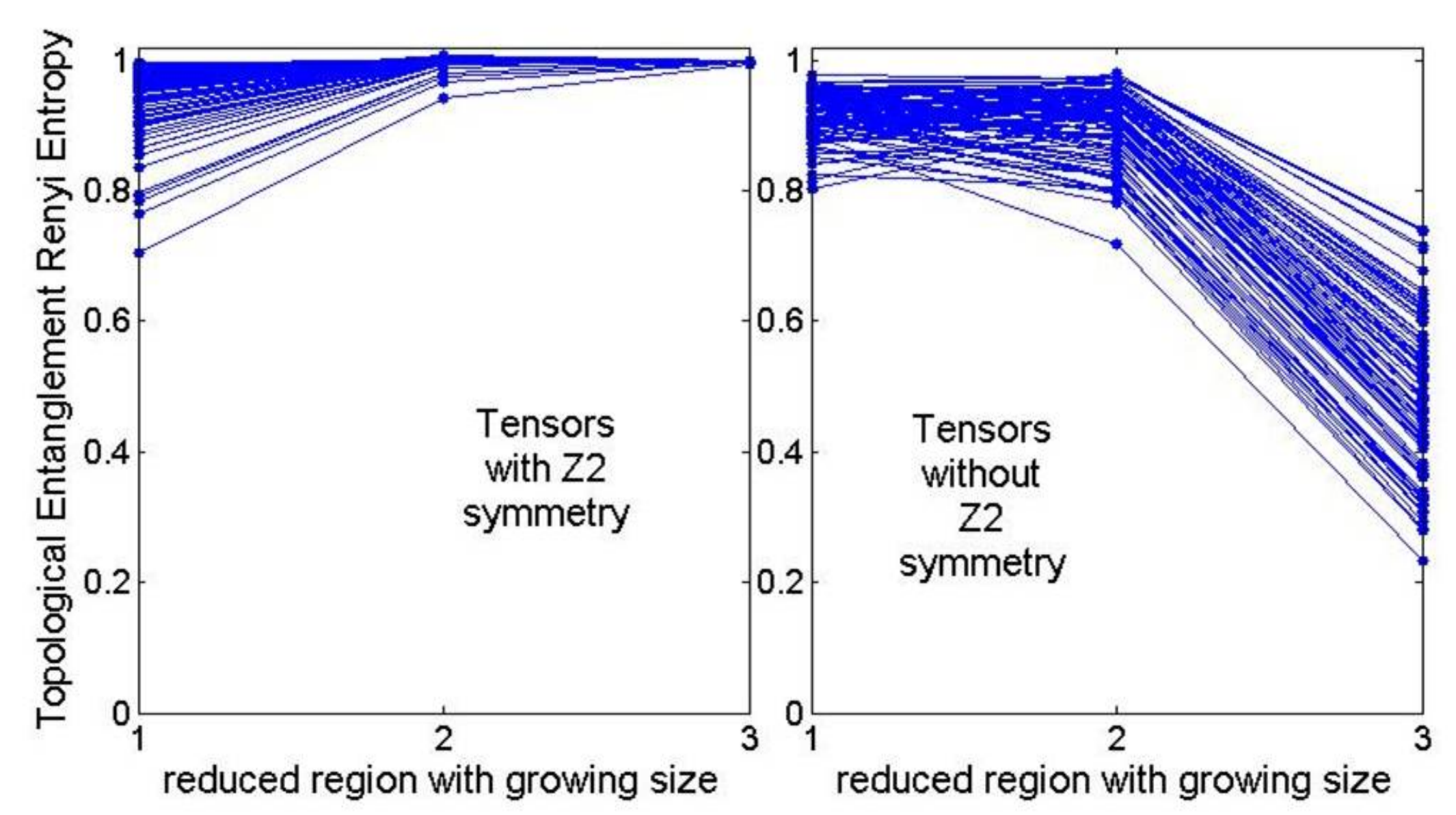} \caption{Topological
entanglement Renyi entropy ($S_{tpr}$) calculated for three
blue regions as shown in Fig.\ref{fig:hex}. Size of regions
grow from region $1$ to $3$. The calculation is done for
$200$ random tensors in the neighborhood of $\mathbb{Z}_2$
tensor. $100$ of them have $\mathbb{Z}_2$ symmetry (plotted
on the left hand side), while the other $100$ have
not (plotted on the right hand side). For tensors with
$\mathbb{Z}_2$ symmetry, $S_{tpr}$ approach $1$ very quickly
as we include more and more qubits in the reduced region,
while for tensors without $\mathbb{Z}_2$ symmetry, $S_{tpr}$
drops towards $0$ as the region gets larger.}
\label{fig:Stpr}
\end{figure}

The necessity of $\mathbb{Z}_2$ symmetry in the generic case is clearly reflected in the following calculation. 
We randomly pick tensors in the
neighborhood of the ideal $\mathbb{Z}_2$ tensor and find the
topological order of the corresponding state numerically. To
do this, we make use of a generalization of topological
entanglement entropy, the Topological Entanglement Renyi
Entropy\cite{Stpr}. Renyi entropy for a reduced density
matrix $\rho$ of order $\alpha$, where $\alpha \geq 0$
\begin{equation}
S_{\alpha}(\rho)=\frac{1}{1-\alpha}log[Tr(\rho^{\alpha})]
\label{Stpr}
\end{equation}
is a valid measure of entanglement. In the limit of $\alpha
\to 1$, it reduces to the usual von Neumann entropy. It was
shown in Ref. \onlinecite{Stpr} that we can replace von Neumann entropy
with Renyi entropy in the definition of topological
entanglement entropy (Eq.\ref{Stp}) and still have a valid
characterization of topological order. The resulting
quantity, topological entanglement Renyi entropy $S_{tpr}$,
does not depend on $\alpha$. We are hence free to choose
$\alpha$ for the ease of computation and we take it to be
$2$. The calculation of Renyi entropy is mapped to the
contraction of a two-dimensional tensor network which can be
computed approximately using the tensor entanglement
renormalization algorithm\cite{Nave}. We take the same
geometry of regions as in Fig.\ref{fig:hex} and the Renyi
entropies of different regions are then combined in the same
way as in Eq.\ref{Stp} to yield $S_{tpr}$. The details of
the computation will be described in Appendix C. Here we
present our result. We restrict ourselves to a small
neighborhood near the $\mathbb{Z}_2$ tensor \begin{equation}
\left
|T_{[n_1n_2n_3]}(i_1i_2i_3)-T_{\mathbb{Z}_2[n_1n_2n_3]}(i_1i_2i_3)\right
|<0.1
\end{equation}
We pick 100 tensors with $\mathbb{Z}_2$ symmetry and plot
how their topological entanglement Renyi entropy scales with
reduced region size in the left half of Fig.\ref{fig:Stpr}
and do the same for 100 tensors without $\mathbb{Z}_2$
symmetry in the right half of Fig.\ref{fig:Stpr}. We see
that for tensors with $\mathbb{Z}_2$ symmetry, $S_{tpr}$
approach $1$ very quickly as we include more and more qubits
in the reduced region, while for tensors without
$\mathbb{Z}_2$ symmetry, $S_{tpr}$ drops towards $0$ as the
region gets larger. This confirms our statement that
$\mathbb{Z}_2$ symmetry is a necessary condition for any generic 
variation of $\mathbb{Z}_2$ tensor to correspond to physical
perturbations of the Hamiltonian and hence characterize
variations within the topological ordered phase. 
The plot also suggests that $\mathbb{Z}_2$ symmetry might be a sufficient condition also.

\section{Conclusion and Discussion}

Our result on $\mathbb{Z}_2$ topological order provides useful perspective on the general relation between tensor variation and Hamiltonian perturbation. First, it is shown that not all variations in tensor correspond to perturbations to the Hamiltonian. For the $\mathbb{Z}_2$ model in particular, based on our calculation of topological entanglement (Renyi)
entropy for tensors in the neighborhood of the ideal
$\mathbb{Z}_2$ tensor [see eq. (\ref{TPS_Z2})], we show that one necessary condition is that, the tensor is invariant
under $\mathbb{Z}_2$ symmetry operation $Z \otimes Z \otimes
Z$ (or any local unitary equivalent operator) on its inner
indices. A generic variation which breaks this symmetry
cannot be induced by local perturbation of the Hamiltonian
and the tensors no longer represent state with $\mathbb{Z}_2$
topological order. This
gives partial answer to the question of what kind of
variations in the $\mathbb{Z}_2$ tensor correspond to
physical perturbations to the Hamiltonian and hence
represent states within the same topological ordered phase. Note that we start with a particular Hamiltonian
in order to better explain the property of the state. 
Our result doesn't depend on this particular form of this Hamiltonian 
and remains valid for any local Hamiltonian of the $\mathbb{Z}_2$ topological ordered state.
(Certain uniformity condition of the Hamiltonian must be satisfied, as pointed out in Ref. \onlinecite{Hastings_TO})
Moreover for simplicity of calculation,
we restricted ourselves to hexagonal lattice in the above
discussion. However, the $\mathbb{Z}_2$ symmetry requirement
is generally true for any lattice structure and the symmetry operation would take the form $Z \otimes Z \otimes...\otimes Z$ on all inner indices(or any local unitary equivalent operator). We expect that
similar necessary symmetry condition also holds for other
quantum double model with gauge symmetry \cite{Qdouble}.
The generalization to other gauge symmetries are discussed 
in more detail in Appendix D.

This understanding will provide important guidance for
variational studies of topological order using tensor
product states. Suppose that, for example, we want to find a tensor 
product state which is the approximate ground state of 
a Hamiltonian with $\mathbb{Z}_2$ topological order.
It is then very important to search within the set of 
variational tensors that have $\mathbb{Z}_2$ symmetry.
If the numerical calculation does not carefully preserve 
this symmetry, we might result in a tensor without 
$\mathbb{Z}_2$ invariance. As the $\mathbb{Z}_2$ breaking 
term can be arbitrarily small, the corresponding tensor 
product state might still give good approximation to 
local properties such as energy, but will have totally 
wrong global properties such as topological order. Then any attempt
to decide the phase diagram based on the state 
would be misleading.

Finally we would like to note that the symmetry
conditions might not be sufficient. A complete understanding
of the correspondence between Hamiltonian perturbation and
tensor variation would be very much desired as it might lead
to full classification of quantum states and quantum phases
using the tensor language. 

Acknowledgement: XC would like to thank Sergey Bravyi and Frank Verstraete for helpful discussions.
BZ is supported by NSERC and QuantumWorks.
ZCG is supported in part by the NSF Grant No. NSFPHY05-51164.
XGW is supported by NSF Grant No. DMR-0706078. 

\section*{Appendix A: Calculating $S_{tp}$ for
$\mathbb{Z}_2$ model with string tension} In this section,
we give detailed procedure of how topological entanglement
entropy $S_{tp}$ of $\mathbb{Z}_2$ model with string tension
can be calculated using the matchgate tensor technique.
Following the definition in \cite{Stp_Kitaev}, we take out a
region (as in Fig \ref{fig:hex}) from the hexagonal lattice
by breaking the $m$ out-going links in half. Due to the
closed loop constraint on the wave function, the boundary
qubits have only $2^{m-1}$ possible configurations $c_i$.
Regrouping terms in the wave function according to different
boundary configurations, we have (up to normalization)
\begin{equation}
|\Phi^g_{\mathbb{Z}_2}\rangle = \sum_{c_i} \alpha_i |\phi^{out}_{c_i}\rangle |\phi^{in}_{c_i}\rangle
\end{equation}
This wave function is automatically in Schmidt-decomposition
form because for different boundary configurations $c_i$,
$|\phi^{out}_{c_i}\rangle$'s are orthogonal to each other,
and so are $|\phi^{in}_{c_i}\rangle$'s. Knowing the norm and
all the $\alpha_i$'s would enable us to calculate entropy of
the reduced density matrix of the the region.

Define rank three tensors $\mathbb{T}$, $\mathbb{T}_0$,
$\mathbb{T}_1$ with inner dimension two as \begin{equation}
\begin{array}{llll} \mathbb{T}(000)=g^2 & \mathbb{T}(011)=1
& \mathbb{T}(101)=1 & \mathbb{T}(110)=1 \\
\mathbb{T}_0(000)=g^2 & \mathbb{T}_0(011)=1 \\
\mathbb{T}_1(101)=1 & \mathbb{T}_1(110)=1 \\ \text{all
others are $0$}
\end{array}
\label{DT}
\end{equation}
It can be verified that the contraction of $\mathbb{T}$ on
all vertices of the hexagonal lattice gives the norm of
$|\Phi^g_{\mathbb{Z}_2}\rangle$. To calculate $\alpha_i$ for
a particular boundary condition $c_i$, replace tensors at
the boundary with $\mathbb{T}_0$ if the boundary qubit is
$0$ and with $\mathbb{T}_1$ if the qubit is $1$ and make
sure the first inner index is on the boundary link.
Contraction of the new tensor network will give
$|\alpha_i|^2$. These three tensors satisfy the conditions
as defined in Ref. \onlinecite{matchgate} and are called `matchgate'
tensors. The contraction of a tensor network of $N$
`matchgate' tensors can be done efficiently (in time
~$N^3$). Therefore, for a fixed reduced region with boundary
length $m$ in a system of total size $N$, the computation of
entanglement entropy takes time polynomial in $N$ but
exponential in $m$.

We start from a small reduced region (dard blue region in Fig
\ref{fig:hex}) with a small $m$, calculate $S_{tp}$ and
increase the total system size $N$ until the change in
$S_{tp}$ is negligible ($<0.01$). We repeat this process for
different values of $g$ and for progressively larger reduced
regions (lighter blue in Fig \ref{fig:hex}). The result is
plotted in Fig \ref{fig:Stp_g}.

\section*{Appendix B: Calculating $S_{tp}$ for
$\mathbb{Z}_2$ model with end of strings} Now we show how
the calculation of $S_{tp}$ can be carried out for
$\mathbb{Z}_2$ model with end of strings, analytically. We
start again with the division of the lattice into sections
$A$, $B$, $C$ as in Fig \ref{fig:hex}. Without the closed
loop constraint, a region with $m$ boundary links has $2^m$
different boundary configurations. Rewriting the
wave function according to different boundary configurations
$c_i$ as \begin{equation}
|\Phi^{\epsilon}_{\mathbb{Z}_2}\rangle = \sum_{c_i} \beta_i
|\phi^{out}_{c_i}\rangle |\phi^{in}_{c_i}\rangle
\end{equation}
we have obtained the Schmidt-decomposed form of the
wave function and all we need to know to calculate entropy
are the $\beta_i$'s and the norm.

Define rank three tensors $\mathbb{S}$, $\mathbb{S}_0$,
$\mathbb{S}_1$ with inner dimension two as \begin{equation}
\begin{array}{llll} \mathbb{S}(000)=1 & \mathbb{S}(011)=1 &
\mathbb{S}(101)=1 & \mathbb{S}(110)=1 \\
\mathbb{S}(001)=\epsilon^2 & \mathbb{S}(010)=\epsilon^2 &
\mathbb{S}(100)=\epsilon^2 & \mathbb{S}(111)=\epsilon^2 \\
\mathbb{S}_0(000)=1 & \mathbb{S}_0(011)=1 &
\mathbb{S}_0(001)=\epsilon^2 & \mathbb{S}_0(010)=\epsilon^2
\\ \mathbb{S}_1(101)=1 & \mathbb{S}_1(110)=1 &
\mathbb{S}_1(100)=\epsilon^2 & \mathbb{S}_1(111)=\epsilon^2
\\ \text{all others are $0$}
\end{array}
\label{DS}
\end{equation}
Contraction of tensor $\mathbb{S}$ on every vertex of the
lattice gives the norm of
$|\Phi^{\epsilon}_{\mathbb{Z}_2}\rangle$. To calculate
$\beta_i$ for a particular boundary condition $c_i$, replace
tensors at the boundary with $\mathbb{S}_0$ if the boundary
qubit is $0$ and with $\mathbb{S}_1$ if the qubit is $1$ and
make sure the first inner index is on the boundary link.
Contraction of the new tensor network will give
$|\beta_i|^2$. The contraction of these two-dimensional
tensor networks can be made efficient by applying a Hadamard
transformation ($|0\rangle \to
(|0\rangle+|1\rangle)/\sqrt{2}$, $|1\rangle \to
(|0\rangle-|1\rangle)/\sqrt{2}$)to each of the three inner
indices of the tensors and transforming them into
\begin{equation} \begin{array}{ll}
\mathbb{S}'(000)=\sqrt{2}(1+\epsilon^2) &
\mathbb{S}'(111)=\sqrt{2}(1-\epsilon^2) \\
\mathbb{S}'_0(000)=1+\epsilon^2 &
\mathbb{S}'_0(100)=1+\epsilon^2 \\
\mathbb{S}'_0(011)=1-\epsilon^2 &
\mathbb{S}'_0(111)=1-\epsilon^2 \\
\mathbb{S}'_1(000)=1+\epsilon^2 &
\mathbb{S}'_1(100)=-1-\epsilon^2 \\
\mathbb{S}'_1(011)=-1+\epsilon^2 &
\mathbb{S}'_1(111)=1-\epsilon^2 \\ \text{all others are $0$}
\end{array}
\label{DS'}
\end{equation}
It is easy to see that the contraction value of this tensor
network can be computed analytically, from which we know
that the entropy of any region with $m$ outgoing links is
\begin{eqnarray} S = m -
\frac{1}{2}\frac{(1+b^{N_i})(1+b^{N_o})}{1+b^{N}}*\ln\frac{(1+b^{N_i})(1+b^{N_o})}{1+b^{N}}
\nonumber \\ -
\frac{1}{2}\frac{(1-b^{N_i})(1-b^{N_o})}{1+b^{N}}*\ln\frac{(1-b^{N_i})(1-b^{N_o})}{1+b^{N}}
\label{S}
\end{eqnarray}
where $b = (1-\epsilon^2)/(1+\epsilon^2)$ and $N_i$($N_o$)
is the number of vertices inside (outside) the region.
$N=N_i+N_o$ is the total system size.

Combining the entropy of different regions according to Eq.
\ref{Stp} and taking the limit $N_i \to \infty$, $N \to
\infty$, we get $S^{\epsilon}_{tp} = 0$ whenever $\epsilon
\neq 0$ for $\mathbb{Z}_2$ model with end of strings.

\section*{Appendix C: Calculating $S_{tpr}$ for random
tensors in the neighborhood of $\mathbb{Z}_2$} Redefining
topological entanglement entropy in terms of Renyi entropy
might simplify the calculation. Specifically, the
calculation of Renyi entropy at $\alpha=2$ for a tensor
product state can be mapped to the contraction of a single
tensor network, which can be computed efficiently in one
dimension and approximated in two or higher dimension.
\begin{figure}[htp] \centering
\includegraphics[width=2.5in]{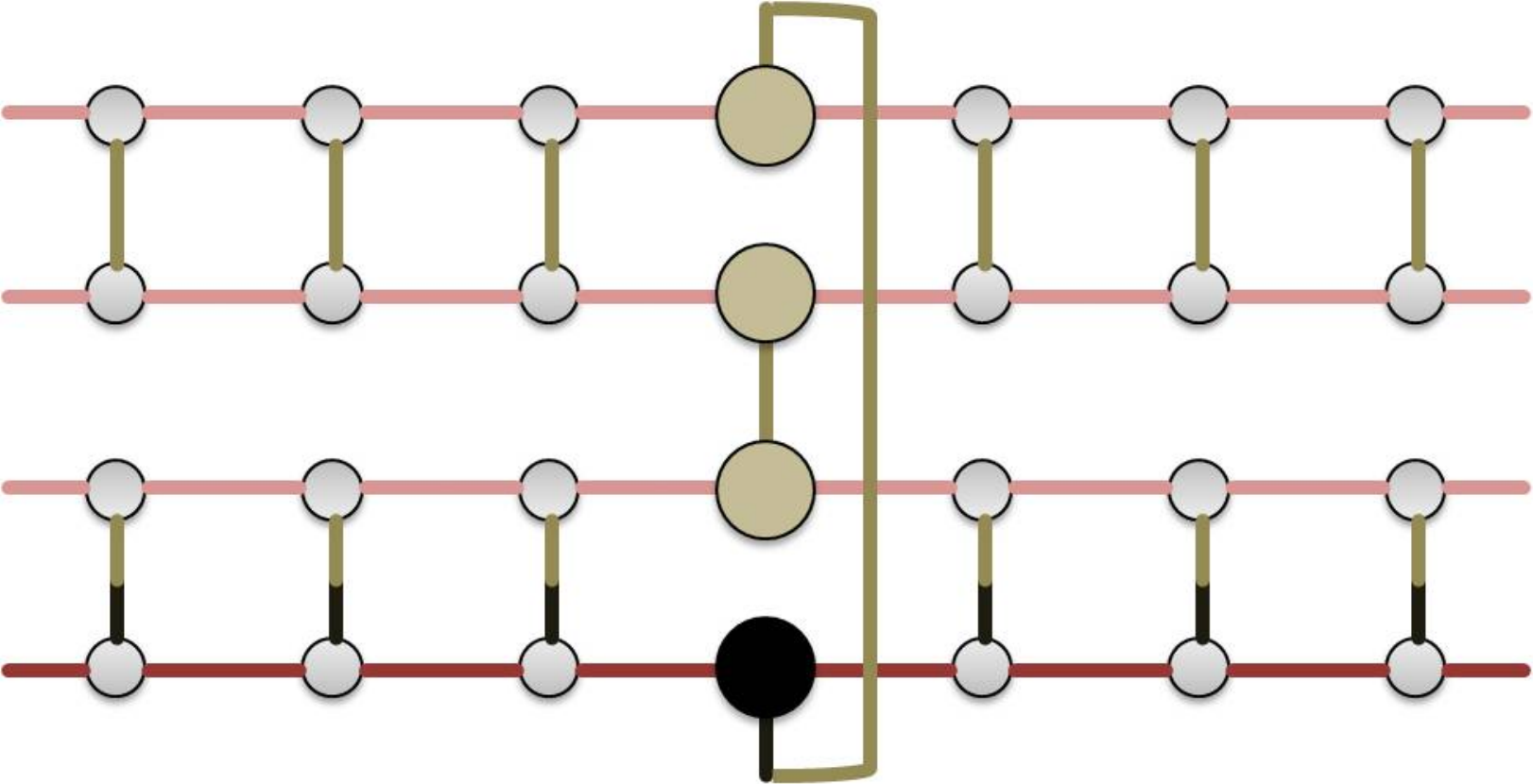} \caption{Tensor
network for calculating Renyi entropy ($\alpha=2$) of the big
site in a one-dimensional tensor product state. The
contraction of the tensor network gives
$Tr(\rho^2)=exp(-S_{2}(\rho))$, where $\rho$ is the reduced
density matrix . The lowest level represents the state,
where red links represent inner indices along the
one-dimensional chain and gray links represent physical
indices. Four copies of the state are stacked together and
corresponding physical indices are connected between the
levels. For physical indices outside the reduced region, the
connection is between levels $1 \& 2$ and levels $3 \& 4$.
For those in the reduced region, the connection is between
levels $1 \& 4$ and levels $2 \& 3$.} \label{fig:Sr2}
\end{figure}
For example, consider a one dimensional tensor product
state (also called a matrix product state). The lowest dark
level in Fig.\ref{fig:Sr2} gives a side view of the state,
where red links represent inner indices along the
one-dimensional chain and gray links represent physical
indices. Renyi entropy at $\alpha=2$ is defined as
$S_{2}(\rho)=-log[Tr(\rho^2)]$. To find out $Tr(\rho^2)$, we
stack four copies of the states together as in
Fig.\ref{fig:Sr2}, connect corresponding physical indices
outside the reduced region between levels $1 \& 2$, $3 \& 4$
and connect those within the reduced region between levels
$1 \& 4$, $2 \& 3$. Contraction of this four-layer
one-dimensional tensor network gives $Tr(\rho^2)$. For two
dimensional tensor product states, the generalization is
straightforward. The only difference is that now we have to
contract a four-layer two-dimensional tensor network. To
this end, we apply the Tensor Entanglement Renormalization
Algorithm \cite{Nave}. Having obtained the Renyi entropy for
different regions, we then combine them to get $S_{tpr}$.

\section*{Appendix D: Gauge symmetry of tensor product states and topological order}

This section discusses in general the relation between gauge
symmetries of tensor product states and topological order,
and the implication of our result on other topological
ordered models with gauge symmetry.

For a tensor product state, the network of tensors which
represents the same state is not unique. In particular, if
we change a pair of connected tensors by rotating the basis
of one of the connected inner index with an invertible
operator $A$ and rotating the other connected inner index
with operator $A^{-1}$, any tensor trace would remain
unchanged and hence the tensor product state remains the
same. This corresponds to inserting a pair of invertible
operators $A$, $A^{-1}$ onto any link in the graphical
representation of the state. Following the definition in
Ref. \onlinecite{SW}, this is called a gauge transformations
of the tensor product state, which form a very large
group. Hence the correspondence between the tensor network
and the physical state is many-to-one.  As a result, the
variation energy as a function of tensors has a very large
symmetry: the variation energy is invariant under the
gauge transformations.

On the
other hand, when we try to find the best description of
ground state for a model Hamiltonian by minimizing energy
with respect to the variations of the tensors, the tensors
that minimize the average energy may not be invariant under
all the gauge transformations and in general have much less symmetry. For
example, in the ideal $\mathbb{Z}_2$ case, the tensors are
only invariant if we insert $Z$, $Z^{-1}$ to all the links
in the 2-dimensional graph. Generalizing this to any
symmetry group and to any dimension $d$, we define the
$d$-dimensional Invariant Gauge Group($d$-IGG). The $d$-IGG
is nothing but the invariant group of the tensors
under gauge transformations. Thus the
minimization of the average energy leads to a spontaneous
symmetry breaking. The $d$-IGG's are the unbroken symmetry
of the tensors that describe the ground state. As we change
the Hamiltonian, the tensors that minimize the average energy
may have some different symmetry structures described by
different $d$-IGG's. As is shown in Ref. \onlinecite{SW}, when $d$ equals the dimension of system space $d_{space}$,
$d_{space}$-IGG (such as the $\mathbb{Z}_2$
symmetry discussed in this paper) can be used to determine the topological orders of a
tensor product state. A closely related concept is discussed in Ref. \onlinecite{Schuch_TO}. Therefore, a change in $d_{space}$-IGG will in
general represent a change in topological order. Apart from $d_{space}$-IGG, the tensors might also have lower dimensional IGG's. For example, if we trivially map every inner index $i$ to $ii$($i=0,1$) in the $\mathbb{Z}_2$ tensor, the tensors still represent the same state but have a $0$-IGG $ZZ$ in additional to its $2$-IGG. However, we believe that such lower dimensional IGG's are not related to the topological order in two dimension and breaking them will not lead to a change in topological order. 

Note that in order to use $d_{sapce}$-IGG of a tensor network to decide topological order, we only require that the network is composed of patches of tensors which are invariant under certain gauge transformations. It is not necessary that every single tensor is $d_{sapce}$-IGG invariant. However, in the generic case, if the single tensors do not have special symmetry structure, it is not possible to have $d_{space}$-IGG invariance on a bigger patch. As discussed in Ref. \onlinecite{PEPS_H}, such a tensor network will generically satisfy a condition called `injectivity', i.e. for a large enough region in the network, when the single tensors are contracted together to form a new tensor, the set of tensor vectors labeled by their physical indices will span the full tensor space of the $n$ outgoing inner indices of the region. Therefore, the tensor network cannot have nontrivial $d_{space}$-IGG. In order for a bigger patch in the network to have $d_{space}$-IGG invariance, it is in general necessary for every tensor to be $d_{space}$-IGG invariant.

Hence, we believe that the invariance of every tensor under $d_{space}$-IGG is a more
general necessary conditions for generic variations of the tensor to
correspond to physical perturbations of the Hamiltonian.
Breaking of the $d_{space}$-IGG invariance of the tensors will in general
correspond to a change in topological order. 
Therefore in a
numerical variational calculation it is very important to
preserve the $d_{space}$-IGG invariance. Otherwise we would not be
able to correctly determine the topological order of the
resulting state from the tensors.

\end{document}